\documentclass[aps, amsmath, amssymb, aps, pre, twocolumn,groupedaddress,superscriptaddress,nofootinbib,longbibliography]{revtex4-1}

\usepackage{multirow}
\usepackage{enumitem}
\setlist{noitemsep,topsep=0pt,parsep=0pt,partopsep=0pt}
\usepackage{array}
\usepackage[english]{babel}
\usepackage[colorlinks,linkcolor=black,citecolor=blue,urlcolor=black]{hyperref}
\usepackage{amsmath}
\usepackage{tabularx}

\usepackage{graphics}
\usepackage{graphicx}
\usepackage{xcolor}
\usepackage[config, labelfont={sf,bf}, textfont=sf]{subfig}
\graphicspath{{}}
\usepackage[suffix=, outdir=./figs/]{epstopdf}
\usepackage{hhline}

\usepackage{lineno}

\modulolinenumbers[5]

\DeclareMathOperator{\LEs}{LE}

\begin{document}

\title{A note on finite-time Lyapunov dimension of the Rossler attractor}

\author{N. V. Kuznetsov}
\email[]{Corresponding author: nkuznetsov239@gmail.com}
\affiliation{Faculty of Mathematics and Mechanics, St. Petersburg State University,
Peterhof, St. Petersburg, Russia}
\affiliation{Department of Mathematical Information Technology,
University of Jyv\"{a}skyl\"{a}, Jyv\"{a}skyl\"{a}, Finland}
\author{T. N. Mokaev}
\affiliation{Faculty of Mathematics and Mechanics, St. Petersburg State University,
Peterhof, St. Petersburg, Russia}

\date{\today}


\begin{abstract}
For the R\"{o}ssler system we verify Eden's conjecture on the maximum of local Lyapunov dimension.
We compute numerically
finite-time local Lyapunov dimensions on the R\"{o}ssler attractor
and embedded unstable periodic orbits.
The UPO computation is done by Pyragas time-delay feedback control technique.
\end{abstract}

\maketitle

\section{R\"{o}ssler attractor and Pyragas stabilization of embedded unstable periodic orbits}

Consider the following R\"{o}ssler system \cite{Rossler-1976}
\begin{equation}\label{eq:rossler}
\begin{aligned}
 &\dot{x} = - y - z, \\
 & \dot{y} = x + a y, \\
 & \dot{z} = b - c z + x z,
\end{aligned}
\end{equation}
with arbitrary real parameters $a,b,c \in \mathbb{R}$.
If $c^2 \geq 4 a b$, then system~\eqref{eq:rossler}
has the following equilibria:
\begin{equation}\label{eq:rossler:equilibria}
O^\pm = (a p^\pm, -p^\pm, p^\pm), \ \ \text{where} \ \ p^\pm = \tfrac{c \pm \sqrt{c^2-4ab}}{2a}.
\end{equation}

For some values of parameters system \eqref{eq:rossler} exhibits chaotic behavior.
To get a visualization of chaotic attractor
one needs to choose an initial point in the basin of attraction of the attractor
and observe how the trajectory, starting from this initial point,
after a transient process visualizes the attractor:
an attractor is called a \emph{self-excited attractor}
if its basin of attraction
intersects with any open neighborhood of an equilibrium,
otherwise, it is called a \emph{hidden attractor} \cite{LeonovKV-2011-PLA,LeonovK-2013-IJBC,LeonovKM-2015-EPJST,Kuznetsov-2016}.
It was discovered numerically by R\"{o}ssler that
in the phase space of system \eqref{eq:rossler}
with parameters $a = 0.2$, $b = 0.2$, $c = 5.7$
there exist a \emph{chaotic attractor} of spiral shape,
which is self-excited with respect to both equilibria $O^\pm$.

\begin{figure*}[ht]
  \centering
  \subfloat[]{
    \label{fig:rossler:upo:all}
  \includegraphics[width=0.4\textwidth]{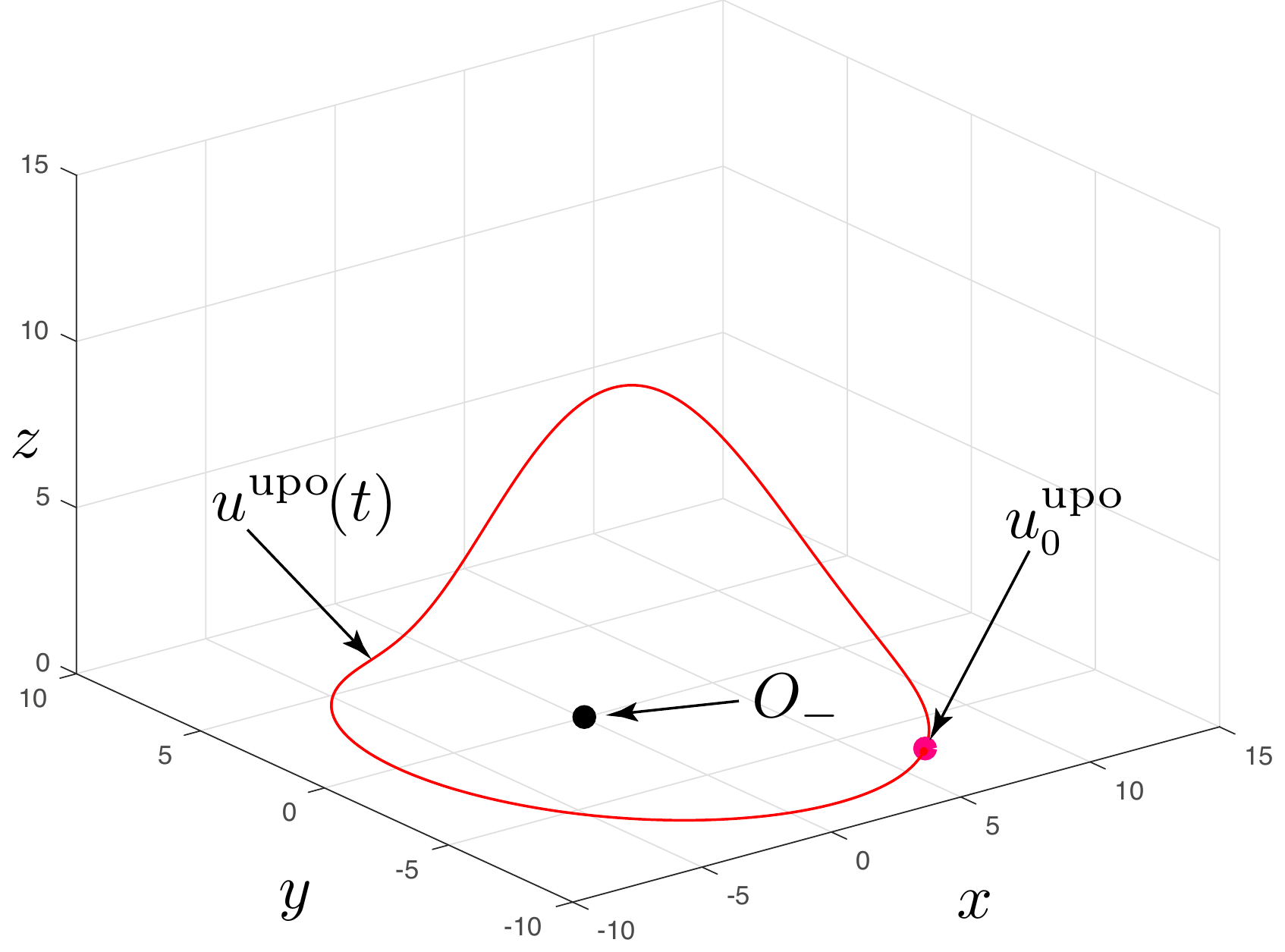}
  }\quad
  \subfloat[]{
    \label{fig:rossler:upo:upo1attr}
    \includegraphics[width=0.4\textwidth]{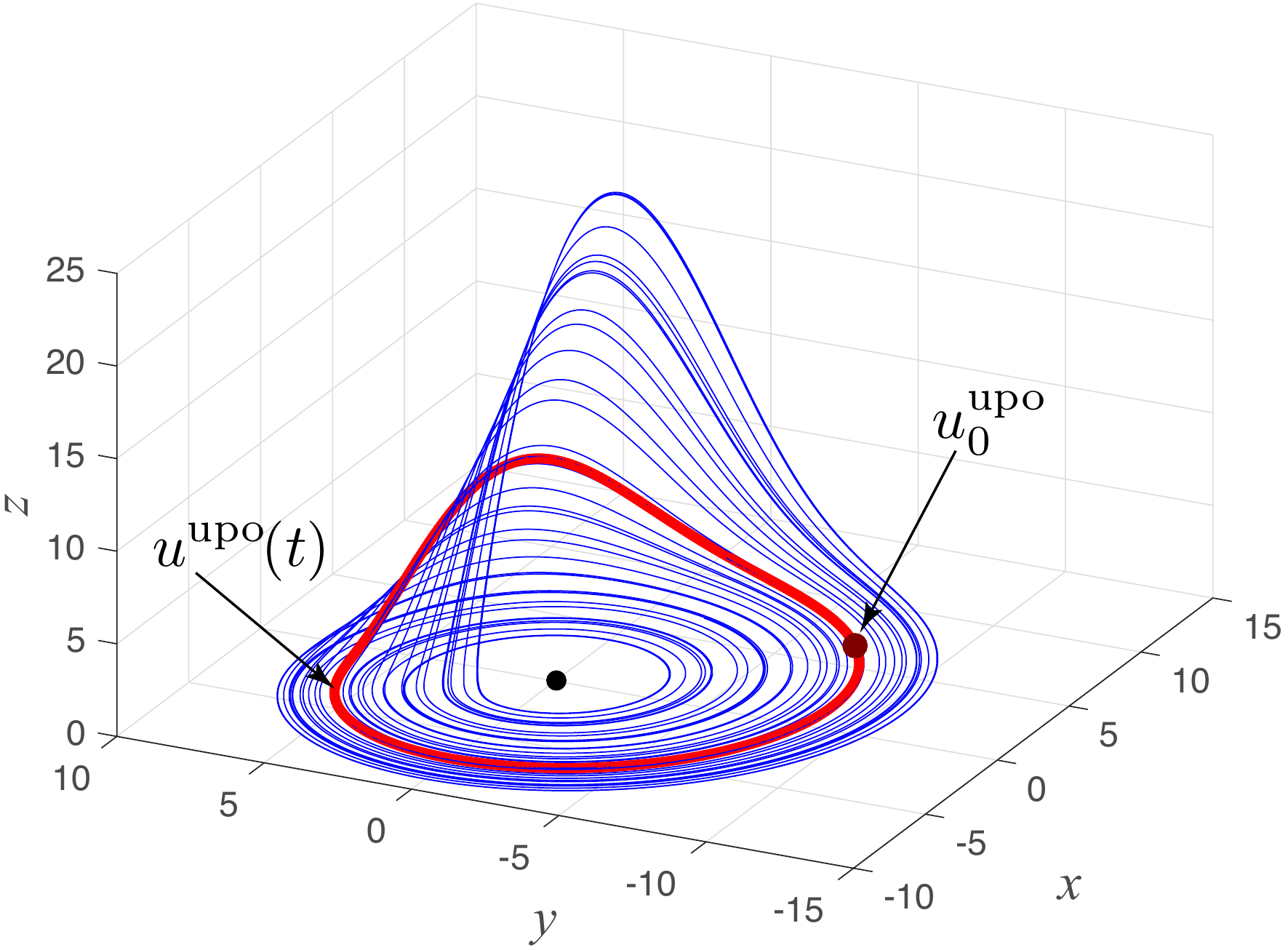}
  }
  \caption{Period-1 (red, period $\tau = 5.8811$) UPO
  in system \eqref{eq:rossler} with parameters $a = 0.2$, $b = 0.2$, $c = 5.7$,
  stabilized using TDFC method.
  }
  \label{fig:rossler:upo}
\end{figure*}

One of the building blocks of chaotic attractor
are embedded unstable periodic orbits (UPOs).
An effective method for the computation of UPOs
is the \emph{time-delay feedback control} (TDFC) approach,
suggested by K. Pyragas \cite{Pyragas-1992}
(see also discussions in \cite{KuznetsovLS-2015-IFAC,ChenY-1999,CruzVillar-2007,LehnertHFGFS-2011}).
Let $u^{\rm upo}(t)$ be an UPO with period $\tau > 0$, $u^{\rm upo}(t - \tau) = u^{\rm upo}(t)$,
satisfying a differential equation
\begin{equation}\label{eq:control_syst}
  \dot{u} = f(u).
\end{equation}
To compute the UPO, we add the TDFC:
\begin{equation}\label{eq:closed_loop_syst}
  \dot{u} = f(u) + k BC^*\big(u(t - T) - u(t)\big),
\end{equation}
where $B,C$ are vectors and $k$ is a real gain.
If $T=\tau$, then $k BC^*\big(u(t - T) - u(t)\big)=0$
along the UPO,
and periodic solution of system \eqref{eq:closed_loop_syst}
coincides with periodic solution of system \eqref{eq:control_syst}.

For the R\"{o}ssler system \eqref{eq:rossler}
we solved numerically system \eqref{eq:closed_loop_syst}
and stabilized a period-1 UPO $u^{\rm upo_1}(t,u_0)$
with period $\tau = 5.8811$
(see Fig.~\ref{fig:rossler:upo}).

Then for the initial point $u^{\rm upo}_0$,
chosen on the UPO $u^{\rm upo} = \big\{u^{\rm upo}(t)$, $t \in [0,\tau]$\big\},
we numerically compute the trajectory
$\tilde{u}(t, u^{\rm upo}_0)$
of system \eqref{eq:closed_loop_syst} without the stabilization (i.e. with $k=0$)
on sufficiently large time interval $[0,T=500]$
(see Fig.~\ref{fig:rossler:upo:upo1attr}).
One can see that on the initial small time interval $[0,T_1 \approx 60]$,
even without the control,
the obtained trajectory $\tilde{u}(t, u^{\rm upo}_0)$
traces approximately the ''true'' periodic orbit $u^{\rm upo}(t, u^{\rm upo}_0)$.
But for $t > T_1$ without control
the trajectory $\tilde{u}(t, u^{\rm upo}_0)$
diverge from $u^{\rm upo}$ and wind on the attractor $\mathcal{A}$.

\section{Finite-time Lyapunov dimension and Eden conjecture}

For an attractor, an interesting question \cite[p.98]{Eden-1989-PhD} (known as Eden conjecture)
is whether the supremum of the local Lyapunov dimensions is achieved on
a stationary point or an unstable periodic orbit embedded in the strange attractor.
In general, a \emph{conjecture on the Lyapunov dimension of self-excited attractor} \cite{Kuznetsov-2016-PLA,KuznetsovLMPS-2018}
is that for a typical system
the Lyapunov dimension of a self-excited attractor
does not exceed the Lyapunov dimension of one of unstable equilibria,
the unstable manifold of which intersects with the basin of attraction
and visualize the attractor.

Below we follow the concept of the
\emph{finite-time Lyapunov dimension} \cite{Kuznetsov-2016-PLA,KuznetsovLMPS-2018},
which is convenient for carrying out numerical experiments with finite time.
The \emph{finite-time local Lyapunov dimension} \cite{Kuznetsov-2016-PLA,KuznetsovLMPS-2018}
can be defined via an analog of the \emph{Kaplan-Yorke formula}
with respect to the set of finite-time Lyapunov exponents:
\begin{multline}\label{lftKY}
   \dim_{\rm L}(t, u)\!=\!d_{\rm L}^{\rm KY}(\{\LEs_i(t,u)\}_{i=1}^3)= \\
   j(t,u) + \tfrac{\LEs_1\!(t,u) + \cdot\cdot + \LEs_{j(t,u)}\!(t,u)
   }{|\LEs_{j(t,u)\!+\!1}(t,u)|},
\end{multline}
where $j(t,u) = \max\{m: \sum_{i=1}^{m}\LEs_i(t,u) \geq 0\}$.
Then the \emph{finite-time Lyapunov dimension}
(of dynamical system generated by \eqref{eq:control_syst}
on compact invariant set $\mathcal{A}$) is defined as
\begin{equation}\label{DOmaptmax}
  \dim_{\rm L}(t, \mathcal{A}) = \sup\limits_{u \in \mathcal{A}} \dim_{\rm L}(t,u).
\end{equation}

The \emph{Douady--Oesterl\'{e} theorem} \cite{DouadyO-1980} implies that
for any fixed $t > 0$
the finite-time Lyapunov dimension, defined by \eqref{DOmaptmax},
is an upper estimate of the Hausdorff dimension:
$\dim_{\rm H} \mathcal{A} \leq \dim_{\rm L}(t, \mathcal{A})$.
The best estimation is called the \emph{Lyapunov dimension} \cite{Kuznetsov-2016-PLA}
\[
   \dim_{\rm L} \mathcal{A}
   = \inf_{t >0}\sup\limits_{u \in K} \dim_{\rm L}(t,u)
   = \liminf_{t \to +\infty}\sup\limits_{u \in K} \dim_{\rm L}(t,u).
\]

\begin{figure*}[t!]
 \centering
 \subfloat[]{
    \label{fig:rossler:upo1:attr}
    \includegraphics[width=0.34\textwidth]{rossler_Chaos_UPO-1.pdf}
  }
 \subfloat[]{
    \label{fig:rossler:upo1:LE1}
    \includegraphics[width=0.33\textwidth]{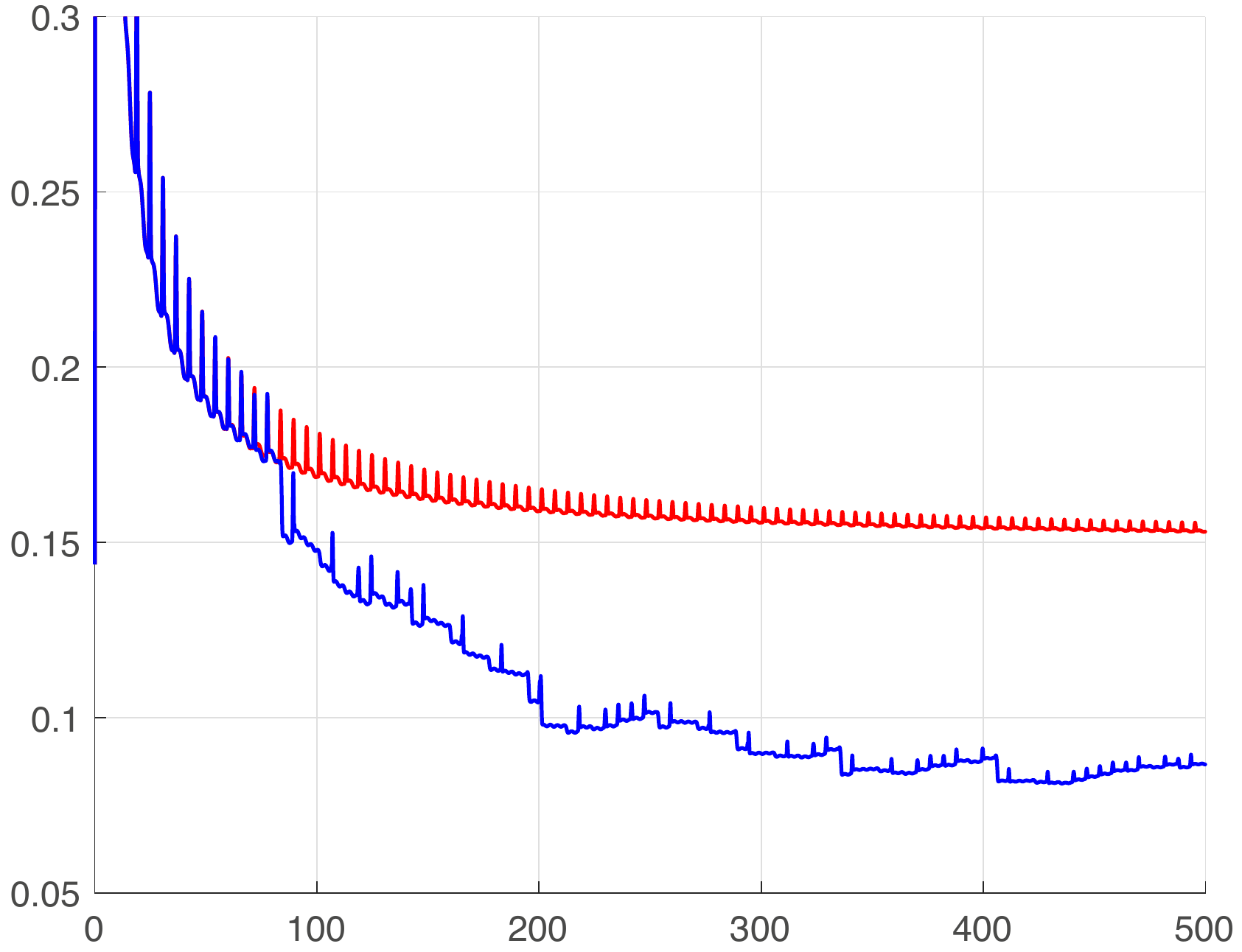}
  }
  \subfloat[]{
    \label{fig:rossler:upo1:LD}
    \includegraphics[width=0.33\textwidth]{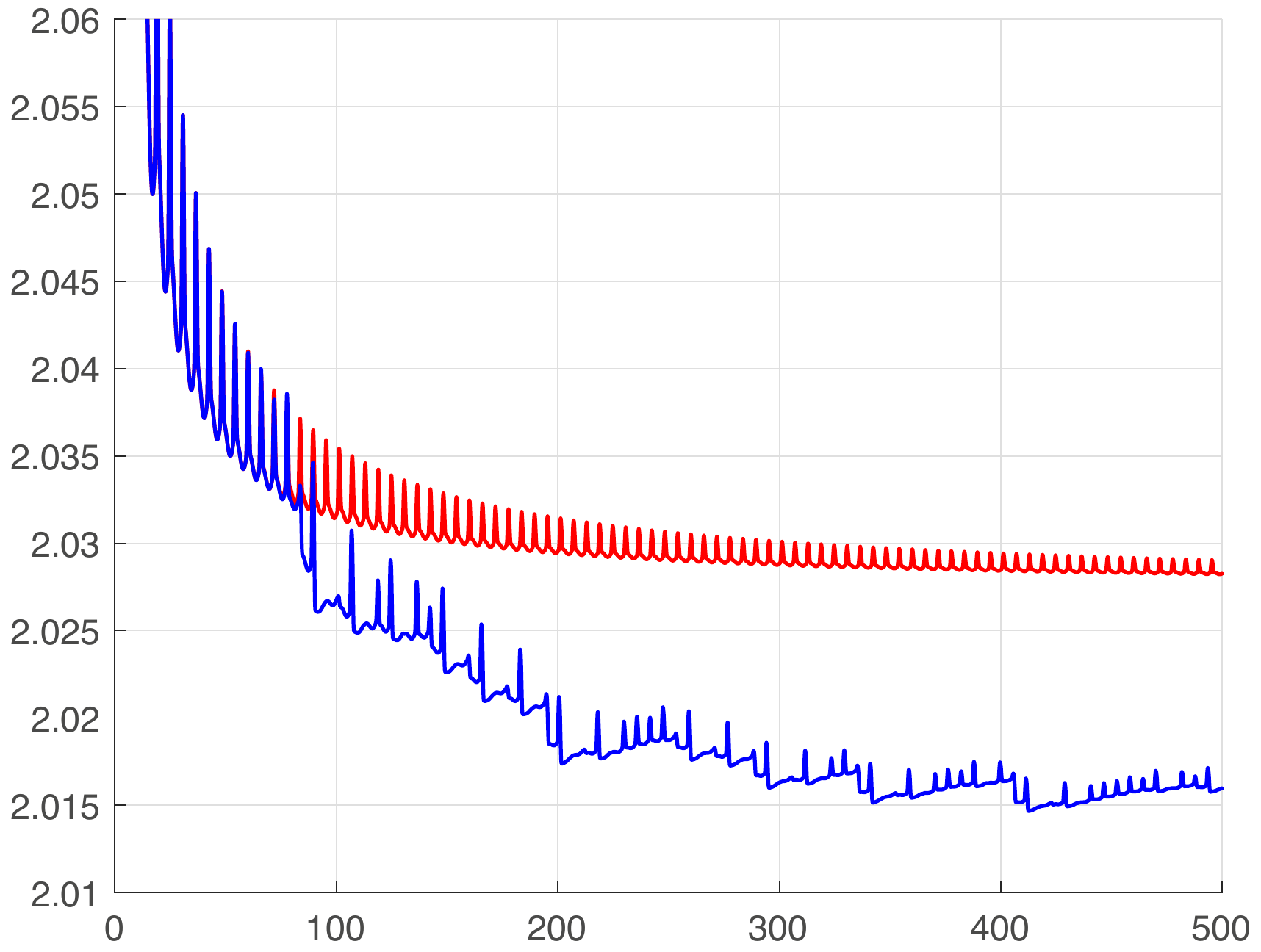}
  }
\caption{$\LEs_1(t,u^{\rm upo}_0)$ and
$\dim_{\rm L}(t, u^{\rm upo}_0)$ on the time interval $t \in [0,500]$
along the UPO $u^{\rm upo}(t)$ (red)
and trajectory integrated without stabilization (blue).
Both trajectories start from the point $u^{\rm upo}_0 = (6.491, -7.0078, 0.1155)$.
}
\label{fig:rossler:upo2}
\end{figure*}

For the R\"{o}ssler attractor
the Lyapunov dimension was estimated as
$2.014$~\cite{FroehlingCFPS-1981},
$2.01$~\cite{SanoS-1985},
$2.0132$~\cite{Sprott-2003,Fuchs-2013},
and $2.09635$~\cite{AwrejcewiczKEDBK-2018});
see also \cite{KuznetsovMV-2014-CNSNS,SprottL-2017}.

Below we use the adaptive algorithm \cite{KuznetsovLMPS-2018} for the computation of
the finite-time Lyapunov dimension and exponents.
We compute: maximum of the finite-time local Lyapunov dimensions
  at the points of grid filling the attractor $\mathcal{A}$, i.e.
 $\max_{u \in C_{\rm grid}} \dim_{\rm L}(t,u)$;
 finite-time Lyapunov dimensions $\dim_{\rm L}(500,\cdot)$
 for the stabilized UPO with periods $\tau = 5.8811$.

The comparison of the obtained values of $\LEs_1(t,u^{\rm upo}_0)$ and
$\dim_{\rm L}(t, u^{\rm upo}_0)$
computed along the stabilized UPO and the trajectory without stabilization
gives us the following results.
On the initial part of the time interval, one can indicate the coincidence of these values
with a sufficiently high accuracy.
For the period-1 UPO and for the unstabilized trajectory
the largest Lyapunov exponents $\LEs_1(t,u^{\rm upo}_0)$
coincide up to the 5th decimal place inclusive on the interval $[0,30.4]$.
After $t > 71.5$ the difference in values becomes significant and the
corresponding graphics diverge
in such a way that the part of the graph corresponding to the unstabilized trajectory
is lower than the part of the graph corresponding to the UPO
(see Fig.~\ref{fig:rossler:upo1:LE1}).

The equilibria $O_\pm$ has simple eigenvalues and, thus,
we have
$\dim_{\rm L} O_+ =  d_{\rm L}^{\rm KY}(\{{\rm Re} \lambda_i(O_+)\}_{i=1}^3) = 3,
  \dim_{\rm L} O_- =  d_{\rm L}^{\rm KY}(\{{\rm Re} \lambda_i(O_-)\}_{i=1}^3) = 2.0341
$.

The period-1 UPO $u^{\rm upo}$ with period $\tau = 5.8811$
has the following multipliers:
$\rho_1 = -2.40398$, $\rho_2 = 1$, $\rho_3 = -1.2946 \cdot 10^{-14}$.
Thus, for the local Lyapunov dimension of the UPO $u^{\rm upo}(t)$
we obtain
$\dim_{\rm L} u^{\rm upo} = d_{\rm L}^{\rm KY}(\{ \frac{1}{\tau} \log \rho_j \}_{j=1}^3) = 2.0274 \lessapprox 2.0283 = \dim_{\rm L}(500,u^{\rm upo})$.

\section{Conclusion}
In this note we have confirmed the Eden conjecture for the R\"{o}ssler system \eqref{eq:rossler}
and obtained the following relations between the Lyapunov dimensions:
\begin{multline*}
  3 \!=\! \dim_{\rm L} O_+ \!>\! 2.0341 \!=\! \dim_{\rm L} O_- \!>\!
  2.0274 = \dim_{\rm L} u^{\rm upo}\\
  > 2.0160 = \max_{u \in C_{\rm grid}} \dim_{\rm L}(500,u)
  \geq \dim_{\rm L} \mathcal{A} \geq \dim_{\rm H} \mathcal{A}.
\end{multline*}

Concerning the time of integration, remark that while the time series obtained from a \emph{physical experiment} are assumed to be reliable on the whole considered time interval,
the time series produced by the integration of
\emph{mathematical dynamical model}
can be reliable on a limited time interval only
due to computational errors
(caused by finite precision arithmetic and numerical integration of ODE).
Thus, in general, the closeness of the real trajectory $u(t,u_0)$
and the corresponding pseudo-trajectory $\tilde u(t,u_0)$
calculated numerically can be guaranteed on a limited short time interval only.
However, for two different long-time pseudo-trajectories
$\tilde u(t,u^1_0)$ and $\tilde u(t,u^2_0)$ visualizing the same attractor,
the corresponding finite-time LEs can be, within the considered error,
similar due to averaging over time
and similar sets of points
$\{\tilde u(t,u^1_0)\}_{t \geq0}$ and $\{\tilde u(t,u^2_0)\}_{t \geq0}$.
At the same time, the corresponding real trajectories $u(t,u^{1,2}_0)$
may have different LEs,
e.g. $u_0$ may correspond to an unstable periodic trajectory $u(t,u_0)$
which is embedded in the attractor and does not allow one to visualize it.

\end{document}